\documentclass{article}
\usepackage[T1]{fontenc}
\usepackage[portuges,english]{babel}
\usepackage{amssymb,latexsym,amsmath,color,graphics,setspace}
\usepackage[colorlinks,linkcolor=blue,urlcolor=blue,citecolor=black,
plainpages=false,pdfpagelabels,breaklinks]{hyperref}
\usepackage{stmaryrd,mathrsfs}

\newtheorem{dfn}{Definition}[section]

\newcommand{\ita}{\textit}
\newcommand{\mcal}{\mathcal}

\newcommand{\lra}{\leftrightarrow}

\title{{Does Newtonian space provide identity to quantum systems?}\thanks{Version Jan 2018.}}

\author{{Décio Krause}\thanks{Partially supported by CNPq.} \\
{\small Department of Philosophy} \\ {\small Federal University of Santa Catarina} \\ {\small Research Group on Logic and Foundations of Science -- CNPq}}
\begin{document}
\maketitle

\hfill{
{

{\small {\sf "We are intrinsically spatial and temporal
beings. We can only observe particular
entities, handle spatial things, and our
thinking depends on the concept of
individuals. How to reconcile nonspatial
quantum characteristics with spatio-
temporal individuation is perhaps the next great problem physics has to overcome."}}

\vspace{2mm} \hfill{\small\sf (S. Auyang, \ita{How is quantum
field possible?}. Princeton Un. Press, 1995, p.165)}

\hspace{5mm}

{\small\sf "In the microworld, we need uniformity of a strong kind: complete indistinguishability."}

\vspace{2mm} \hfill{\small\sf (F. Wilczek and B. Devine, \ita{Longing for Harmonies: Themes and Variations from Modern Physics}. Penguin Books, 1987, p,135)}
}

\medskip
\begin{abstract} \noindent Physics is not just mathematics. This seems trivial, but poses difficult and interesting questions. In this paper we analyse a particular discrepancy between non-relativistic quantum mechanics (QM) and `classical'  (Newtonian) space and time (NST). We also suggest, but not discuss, the case of the relativistic QM. In this work, we are more concerned with the notion of space and its mathematical representation. The mathematics  entails that any two spatially separated objects are necessarily \ita{different}, which implies that they are \ita{discernible} (in classical logic, identity is defined by means of indiscernibility) --- we say that the space is $T_2$, or "Hausdorff". But when enters QM, sometimes the systems need to be taken as \ita{completely indistinguishable}, so that  there is no way to tell which system is which, and this holds even in the case of fermions. But in the NST setting,  it seems that we can always give an \ita{identity} to them by means of their individuation, which seems to be contra the physical situation, where individuation (isolation) does not entail identity (as we argue in this paper). Here we discuss this topic by considering a case study (that of two potentially infinite wells) and conclude that, taking into account the quantum case, that is, when physics enter the discussion, even NST cannot be used to say that the systems do have identity. This case study seems to be relevant for a more detailed discussion on the interplay between physical theories (such as quantum theory) and their underlying mathematics (and logic), in a simple way apparently never realized before. 

\medskip
\noindent Keywords: individuality, identity of quantum particles, spatial identity, space and time in quantum mechanics, underlying logic of physical theories.
\end{abstract}

\section{The problem}
The first notice we have contrasting quantum mechanics (QM) and logic was in the the famous paper by Birkhoff and von Neumann \cite{birvon36}. There, they have claimed that `the logic' of quantum mechanics should not be classical logic, since the distributive law $a \wedge (b \vee c) = (a \wedge b) \vee (a \wedge c)$ would not hold in the quantum domain. Ever since then \ita{quantum logic} became a wider field of knowledge, turning nowdays most to quantum computation and quantum information \cite{dalgiugre04}. For a general view on the subject, see \cite{bub17}, \cite{bokjae10}. 

In this paper I shall present a different way of questioning the coherence of quantum mechanics and its underlying mathematics. It is impressive, at least to me, that this fact was not yet noticed in the literature (as far as I know). It is related with the use QM makes of Newtonian space (and time) described in a mathematics that does not conform with the physical interpretation. So, logic, here, encompasses mathematics (we identity logic as a strong mathematical system such as the ZFC system --- Zermelo-Fraenkel with Choice). Furthermore, I shall refer to QM in a very general (and perhaps imprecise) way, meaning the conjunction of a cluster of `theories' or ways of describing apparently empirically equivalent formalisms; for some of them, see \cite{sty01}. If pressed, we shall say that we are considering QM according to Schrödinger's picture in Hilbert spaces formalism. We use the terms `particle' and `quantum system' interchangeably, with the second being more precise.

Roughly,  our idea can be described as follows.
There is an incompatibility between the quantum underlying space and time and quantum mechanics proper. The mathematical space is \ita{Hausdorff}, which means that any two separated objects represented by \ita{points} in that space must be taken as \ita{distinct} in a very strong sense. They are \ita{different}. From a logical and philosophical point of view, this means that they may conform to a \ita{theory of identity}, here taken as that of ZFC; so, one of them has a property not hold by the other, hence the logic entails their difference by some form of Leibniz's Principle of the Identity of Indiscernibles (if two objects don't share \ita{all} their properties, they are different), which is a theorem of classical logic. The way to see this in a Hausdorff space is jut to take to disjoined open balls centered in the points where the objects are supposed to be located, and bingo! The property if precisely that one whose extension is one of the open balls. 

But \ita{physical quantum mechanical space}, whatever it means, is not Hausdorff. Two entangled quantum systems cannot be put in complete isolation; nonlocal correlations ever exist. Even by considering just one system confined by strong magnetic fields (traps) as it became common to consider (see below), it cannot be said to be \ita{completely} isolated; we shall exemplify this case by supposing infinite potential wells, which are pure idealizations. The first remark is that h although quantum systems are taken as pontual in the mathematical setting, from the physical point of view they cannot be taken as exactly pontual, that is, as precisely localized objects. As Carlo Rovelli remarks \cite{rov17},\footnote{The reduced Planck constant, $\hbar$ equals $h/2\pi$, where $h$ is Planck's constant.} "the major physical characterization of quantum theory is that the volume of the region $R$ where the [quantum] system happens to be cannot be smaller that $2 \pi \hbar$: $$Vol(R) \geq 2 \pi \hbar\,."$$  Ok, you may say, let us take them as  represented not by points, but as something confined in non-empty open balls in NST, and let us take these balls at a distance greater than de Broglie's wavelength $\lambda = h/p$, where $p$ is the momentum of the particle. But, in such a situation, the most the systems gain is what Dalla Chiara and Toraldo di Francia call \ita{mock individuality} (see \cite{daltor93}), yet and the physicist feel free to speak that there is a particle in the surroundings of the first point and \ita{another} in the vicinity of the second point, so that we can even \ita{name} them momentaneously as Peter and Paul (loc.cit.). But, as they also remark, this identity (they speak on `individuality') has a brief duration; as soon as the particle meets others of the similar species, their "mock" identity is lost forever (we shall say what we understand by identity very soon). Due to the characteristics of QM, as we shall see, it is impossible to keep the particles as completely isolated, so a legitimate identity (in the intuitive sense of a particle being identical just to itself) cannot be even considered.\footnote{Really, by hypothesis and \ita{individual}, something having this intuitive \ita{identity}, is the only having such an identity. But any theory of identity requires a metaphysics of identity to give philosophical support to the idea, as we shall discuss at section (\ref{iii}), and quantum systems have not an (metaphysical) individuality criterion; \ita{whatever} quantum system of the right kind would serve to make the physical predictions; it is not necessary that the system be \ita{that} specific one, \ita{that} positron, say.} 

Mathematically, the above discourse has to be qualified, for we need to make clear the sense of many used concepts, such as "momentaneously", "brief duration" and others referring to time in order to cope the quantum case. But, by considering QM, the open balls referred to above would never be disjoined once complete isolation cannot be reached. Below we take a case study dealing with an extreme situation, that of two infinite potential wells. Our conclusion will reinforce the idea that we need to distinguish between at least between two levels of languages, one purely mathematical, which describes the formalism, and a `philosophical' one, where the physical interpretation is given. In our account, answering Omnès question posed next, these languages do not need to be consistent one another. Let us see the details.

\section{Omnès two languages}
In discussing the idea that physics comprises two languages, Roland Omnès \cite{omn99} addresses that

\begin{quote}
"Physics, being both an empirical and a theoretical science can only be fully expressed by means of two distinct languages, or two different kinds of propositions. There is a mathematical language for theory. In the case of quantum mechanics, the framework of this language is provided by the theory of Hilbert spaces. Some of its main propositions express how wave functions evolve in time; other propositions may state the value of matrix elements for observables, from which one
derives spectra, probabilities and therefrom cross sections and the statistics of measurements.

There is also another language dealing with empirically meaningful propositions. These propositions describe an experimental set-up; they state which events occur; they indicate the reading on a voltmeter or a measuring device. They are concerned directly with experiments, expressing: how these experiments are performed and their results.
The existence of two languages in physics may look trivial, but it raises the question of their relation and of their mutual consistency, which is the backbone of \ita{interpretation}."
\end{quote}

Of course Hilbert spaces are not the only way to erect a quantum mechanics (see \cite{sty01} for \ita{nine} different alternatives), but Omn\`es is right concerning the standard way of considering the quantum formalism. The most important point to our concerns here is the alleged possible \ita{consistency} of these two languages. But, first, let me try to elaborate the distinction in order to appropriately deal with the idea (to which we agree almost \ita{in totum} with Omnès, as the reader can see in \cite{kraare16}). The first language, let us call it $\mcal{L}_T$, is the object language of the theory proper, and here we suppose that the theory is axiomatized. The second, let us call it $\mcal{L}_M$, is the metalanguage, the one which provides, according to Omnès, the empirical meaningful propositions and by means of which we start thinking about the subject; see  \cite{kraare16} again for more discussions on this topic. 

\ita{Consistency} is a term that has an intuitive appeal, and it seems that it is in this sense that it is used in physics, but adquires a precise meaning only inside a formal system whose language comprises a negation symbol `$\neg$'. Thus, we can say that a theory $T$ is consistent in two ways: \ita{syntactical consistency} means that there is no formula $\alpha$ such that $T \vdash \alpha$ ($\alpha$ is deduced in $T$) and $T \vdash \neg\alpha$; \ita{semantic consistency} (stronger that the syntactic version) means that $T$ has a model, that is, there is an interpretation of its non-logical symbols that make the non-logical axioms true or, alternatively, theorems of the metatheory (which without loss of generality can be assumed to be a set theory, such as the system ZFC --- Zermelo-Fraenkel set theory with the axiom of choice. For the differences between these alternatives, see \cite{kraare16}). 

But these definitions apparently are much for Omnès requirement. What he is supposed to require is that there should be no discrepancies between the language of the theory and the language which expresses its empirical propositions. And it will be in this sense that we shall consider the case study of this paper. We take $\mcal{L}_T$ as the standard Hilbert space formalism for non-relativistic quantum mechanics (NQM), and for $\mcal{L}_M$ the metamathematical language of ZFC, where we express the space-time (in special, space) counterpart of this theory (see the Appendix to understand how space and time may enter the quantum formalism). Notwithstanding we consider NQM, the conclusions of this paper can be also extended to the relativistic views, since Minkowskian space is also Hausdorff. Anyway, we shall not make this consideration here.  

\section{NQM and its consequences}\label{NQM}
NQM is grounded on `classical' concepts of space and time. In short, and without going to a precise description (details in  \cite[Chap.17]{pen05}, \cite{ste67}, \cite{wea17}), the space and time setting can be roughly identified with the $\mathbb{E}^4$ Euclidean manifold.\footnote{Trully, the definition is more complex, but this is enough for now.} In this framework, due to its topological characteristics, if we have two distinct objects, they can be located into two disjoined open sets and this provides them an \ita{identity}: we can call Peter the first object and Paul de second one, and if it is Peter who is in the first open set, then Paul is not there, so they present a property not shared by the other and them, by one of the fundamental rules of classical logic, namely, Leibniz's Principle of the Identity of Indiscernibles (PII), they are \textit{different} (not equal, not the same object).\footnote{By the way, this is the meaning of \ita{numerical identity}: in classical settings, numerically identical objects are not \ita{different} objects, but \ita{the very same} one.} Furthermore, if we move Peter, say by an orthogonal transformation or if we translate it, or both,\footnote{The composition of a translation with an orthogonal transformation is called a rigid motion, and they are typical of Euclidean geometry and of classical mechanics.} we can always recognize that it was Peter who has being moved, for we can trace back the motion (the transformations are invertible) and identify the object by the open set it belonged to. To make an analogy which will be useful later, we can say that every object in this manifold (that is, every object \ita{represented} in such a mathematical framework) has an identity cart, a document that enables us to identify the object \ita{as such} in whatever situation or context.  We shall say that it is an \ita{individual}. An individual, thus, is something that \ita{has identity}, in the sense of possessing an identity card. The particular object is the unique one  with such an identity card, and every \ita{other} object presents a distinction from it, for this object has at least one property shared by no other object, namely, its identity. If we admit the existence of \ita{numerically distinct} objects, that is, entities that differ \ita{solo numero}, just by being more than one but without any intrinsic distinction, then they would be indiscernible, \ita{non-individuals}, entities \ita{without identity}. But classical logic, the underlying logic of NQM is Leibnizian: there are no distinct indiscernible objects. Indiscernible objects (objects partaking all their properties) are identical, are the very same object. This is classical logic, this is classical mathematics, this is NST.

In other words, classical logic, standard mathematics (say, that mathematics that can be developed in a set theory such as the Zermelo-Fraenkel (ZF) set theory) and classical mechanics are Leibnizian in this sense. In classical physics, even entities like electrons, yet having the same characteristics, don't lose their identity, for they have different and impenetrable trajectories which serve to identify them in every instant of time. They have that what  Heinz Post has termed \ita{transcendental individuality} (see \cite{pos63} and \cite[p.11]{frekra06}).

Quantum mechanics, it is agreed by most physicists and philosophers, is different. As Landau and Lifshitz have said (adapting), due to the uncertainty principle, if the position of an electron is exactly known in a given instant, its coordinates have no definite values even at the next instant: ``by localizing and numbering the electrons at some instant, we make no progress towards identifying them at subsequent instants; if we localize one of the electrons, at some other instant, at some point of space, we cannot say which of the electrons has arrived at this point." \cite[p.227]{lanlif91} (see also the quotation at the end of this paper).

But wait! How can we not be able to identify the particles if they are represented in classical space and time (NST)? How can we explain this fact? Here we propose a situation that serves for the analysis of this puzzle. The core idea is that 
once we assume that quantum systems are not mere mathematical entities, there cannot be identity (in the standard sense of numerical identity) provided to them by NST. In other terms, when physics enter the scenario and provides situations which are not purely mathematical (and I suppose we all agree that physics \ita{is not} mathematics),\footnote{But we may recall Max Tegmark's suggestion that the universe is a mathematical structure, the \ita{mathematical universe hypothesis}, or MUH (see the Wikipedia entry on \href{goo.gl/GgzKzL}{MUH}). It should be observed, despite this subject is out of the aims of this paper, that this thesis has problems as the following one: \ita{if} MUH is correct, then the mathematical structure needs to be constructed in some mathematical apparatus. Which one? Furthermore, which mathematical structure is the universe? The interested reader can also have a look at Sabine Hosenfelder's review of Tegmark's book \ita{Our Mathematical Universe} in her blog \href{http://backreaction.blogspot.com.br/2017/11/book-review-max-tegmark-our.html}{BackRe(Action)}.} 
we need to take into account the `second language' mentioned by Omnès, namely, the language of physics properly speaking (in distinction from its mathematical language), the situation changes so that we cannot say that the distinctly located quantum systems possesses identity. 

To provide an analysis of the situation, we consider what is perhaps the most limiting case where apparently there would be no doubt about the identity of quantum systems, namely, two infinite potential wells with one particle each, the particles being of the same species (`identical' in the physicists' jargon). By supposing that the particles cannot scape the wells, their positions in space would serve to provide them an identity card, an identity. We shall see that this conclusion cannot be reached so easily in `real' physical situations, so there is an apparent contradiction between that what QM says and that what its underlying mathematics enables us to do (specially `classical' space and time). So, Omnès two languages are, in this case, not mutually consistent. We conclude with foundational considerations of both physics and logic.

\section{Two wells and `classical' identity}\label{idd}
The situation we shall consider is formed by two infinite potential wells.   If the wells are far apart, it is possible for practical purposes (this is quite an important point) to have a well 1-state and a well 2-state in which the considered particles are bound.\footnote{I have said "for practical purposes"\  because the notions of "being close"\ and "being far apart"\  are relative. In reality, from a theoretical point of view, tunneling effects cannot be supposed do not exist, as we shall see below, since the wave functions spread out in all space.}    Let us suppose that the potentials of the wells are given as follows (see Figure 1).

\begin{equation*}
    V(x) = \begin{cases}
               0               & x_1 - \epsilon\ < x  < x_1 + \epsilon\  \mathrm{and}\ x_2 - \epsilon\ < x  < x_2 + \epsilon \\
               \infty              &  \mathrm{otherwise}, \\
           \end{cases}
\end{equation*}

\noindent where  $x_1$ and $x_2$  ($x_1 \not= x_2$) are the centers of the wells (each of length $2\epsilon > 0$) in the $x$ axis. We can separately solve the independent of time Schrödinger equations for the two wells and get the corresponding wave-functions that describe the energies of the particles in the wells, one for each well \cite[p.24ff]{gri95}.
But these wave-functions can also be used for granting us the existence of the two particles at those distinct locations (being $x_1+\epsilon < x_2 - \epsilon$). Due to the topological structure of this $T_2$ space-time manifold, independently of the distance between the wells we can (mathematically) find two disjoined open sets $A$ and $B$ containing the space regions corresponding to the wells, so that we can surely say that the particles belong to disjoined open sets in the manifold. \ita{Thus}, we conclude grounded in classical logic, they are different.

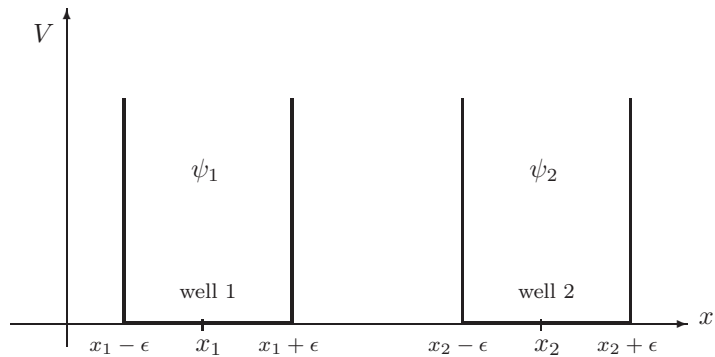
\begin{figure}\label{2wells}
\setlength{\unitlength}{1.5mm} \centering
\begin{picture}(30,30)
\put(-15,5){\vector(1,0){60}}
\put(-10,3){\vector(0,1){30}}
\put(-5,5){\line(0,1){20}}
\put(10,5){\line(0,1){20}}
\put(-4.9,5){\line(0,1){20}}
\put(9.9,5){\line(0,1){20}}

\put(25,5){\line(0,1){20}}
\put(40,5){\line(0,1){20}}
\put(25.1,5){\line(0,1){20}}
\put(39.9,5){\line(0,1){20}}

\put(-5,5.15){\line(1,0){15}}
\put(25,5.15){\line(1,0){15}}
\put(46,5){$x$}
\put(-13,30){$V$}
\put(0,7.3){\footnotesize{well 1}}
\put(30,7.3){\footnotesize{well 2}}
\put(1,18){$\psi_1$}
\put(31,18){$\psi_2$}
\put(2,4.4){\line(0,1){1}}
\put(1.4,2.6){$x_1$}
\put(32,4.4){\line(0,1){1}}
\put(31.4,2.6){$x_2$}
\put(7,2.6){\footnotesize $x_1+\epsilon$}
\put(-8,2.6){\footnotesize $x_1-\epsilon$}
\put(22,2.6){\footnotesize $x_2-\epsilon$}
\put(37,2.6){\footnotesize $x_2+\epsilon$}
\end{picture}
\caption{\footnotesize{Two infinite potential wells. Are they conferring identity to the particles described by $\psi_1$ and $\psi_2$?}}
\end{figure}

But, what does \ita{different} mean?  Let us have a look on the underlying concept of identity before to continue with the discussion about the wells.

\section{Identity, individuality, individuation}\label{iii}
In section (\ref{NQM}), we made reference to an \ita{identity card} for an individual. Such an identity card is provided by some metaphysical principle we call a Principle of Individuality, something which provides to an object (and to the objects in general) a form of ascribing it its  role as an \ita{individual}. Important to distinguish this notion from that of \ita{individuation}, which is an epistemological notion we may have to be able to say that some object is \ita{isolated}, seen as one. Let us be more specific on these points; for more details, see \cite{kraare18}. 

We have an intuitive idea of identity and difference. Identical things are the very same thing; a thing is \ita{identical} justo to itself and with nothing else, so two or more things are \ita{different} things. Thus, more than one thing entails that they are \ita{different}. This belief is grounded on a metaphysical assumption about individuality that goes back at least to the Stoics (cf. \cite[pp.339]{jam66}) but had its fortification within Leibniz's metaphysics. 



The main philosophical question is this: if there is a difference between two objects, where this difference resides? The answers belong to one of the two main metaphysical `theories of individuality', namely, \ita{substratum theories} and \ita{bundle theories}. 
The first  assumes the existence of some kind of substratum underlying the thing's properties. This mysterious substratum receives different names, but with the same main characteristics: \ita{haecceity}, \ita{primitive thisness}, \ita{quid}, etc. which, although presenting differences, shall be taken here as synonymous. So, two things may partake all their properties or characteristics, but differ in what respects their substratum, something that cannot be described by (or reduced to) properties. The problems regarding such a view in quantum mechanics can be seen in \cite{tel98}, the main one being the difficulty of explaining what should be such a substratum, since it cannot be described by means of properties. Bundle theories dispense any kind of substratum; the individuality of a thing (hence, its identity) is given only by its properties, either one property or a collection of them. This view also presents some problems; let us mention one, perhaps the most important one. The question is this: how can we be sure that there are no two or more things partaking the same collection of characteristic properties? The answer is that we don't know; we need to assume this idea or reject it based on metaphysical groundings. Leibnizian metaphysics takes this hypothesis, and it has been incorporated to our preferred pantheon of assumptions. 

Classical logic and standard mathematics are Leibnizian in this sense, which means that the thesis that distinct things have distinct properties is a logical fact, that is, the notion of identity is taken as a logical idea. Really, take a standard set theory (the mathematical framework where NST is developed). Given an object $a$ whatever (described in such a set theory), we can define the `property' of being identical to $a$ as follows: $I_a(x) \lra x \in \{a\}$, by considering that the unitary set of $a$ can be formed in the theory for any $a$.\footnote{Thus result holds even if $a$ is an ur-element, that is, an entity that is not a set but which can be member of sets.} Thus, the only object with this property is $a$ itself, so it presents a property shared with no \ita{other} object. This informally means that any two different entities represented in such a framework have distinct properties: absolute indiscernibility entails identity. Alternatively, we can interpret $I_a$ as "to belong to a suitable open ball centered in $a$" when the topology is Hausdorff. 

More technically, the  Leibnizian notion of identity is \ita{extensional} in the following sense: two sets are identical (they are the very same set) if and only if they have the same elements. Two ur-elements are identical if and only if they belong to the same sets.\footnote{Permutations of ur-elements can be extended to automorphisms of the relevant structures. So, may be we are not able to discern among ur-elements, but this is an epistemological limitation. The above argument involving $I_a$ shows that ur-elements are individuals in the sense explained above.} In other terms, a set, or a context (whatever it is) changes if a thing belonging to it is exchanged by a thing not belonging to it. We can express this idea by the following theorem of extensional set theories (theories encompassing an Axiom of Extensionality):

\begin{equation}\label{extens}
x \in A \wedge y \notin A \to \Big((A - \{x\}) \cup \{y\}) \not= A\Big).
\end{equation}

In quantum mechanics, apparently things don't run this way. Think of ionization. Let $A$ be an atom whatever in its fundamental state (for instance, an Helium atom). We can ionize the atom by realizing an electron and getting a positive ion, $A^+$. After some time, we can make the ion absorbs an electron again, turning to be a neutral atom once more. Questions: is the `new' neutral atom the same as the `old' one? Are the realized electron and the absorbed electron the same electron? Obviously that these questions cannot be answered out of serious conceptual doubts or strong metaphysical assumptions. In fact, we cannot say that we are realizing this or that electron, but just \ita{one} electron; the same concerning the absorbed electron. The intuitive notion of identity seems to be not applicable in this domain. Electrons, atoms and other \ita{quanta} don't have identity cards; there is no Principle of Individuality applicable to them.\footnote{We do not consider Bohm's approach for we regard it as quite obscure, postulating some entities, such as the infamous \ita{quantum potential} or them the position of the particles, which cannot be accessed in a `natural' way --- se also \cite{rov17}.} We believe that we don't have a Principle of Individuality that provides an `alibi' to an electron, to use Hermann Weyl's known passage (see \cite[p.224]{frekra06}, \cite{kraare18}). And, more importantly, they don't acquire identity even after being realized, for once the electron merges the environment, it\footnote{In so far as we can speak of \ita{it} and \ita{others} --- see below how to make language precise.} becomes tangled up with `other' electrons, so that we cannot identify it ever more.

It seems to me, as we have discusses elsewhere (see \cite{arekra14,frekra06}),  that the better and fair idea is to say that the neutral atoms are \ita{indiscernible} or \ita{indistinguishable} from one another (see below), so as are the two electrons. Physicists surely will agree, saying that this is the obvious conclusion to be made. But we remark that this is not \ita{so} obvious as it seems: in classical logic (set theory included), indiscernibility entails identity, for the very notion of identity is introduced by means of indiscernibility (Leibniz's Law): two things are identical if and only if they share all their properties. But, in the permutation of quantum entities, what we are looking for is something like the equivalence indicated bellow, where $\equiv$ stands for a relation of indiscernibility and $x$ and $y$ stand for the realized and the absorbed electrons and $A$ is the neutral atom:

\begin{equation}\label{indist}
(A - \{x\}) \cup \{y\}) \equiv A,
\end{equation}

\noindent but we neither have true conditions for asserting that electron $x$ belongs to the atom nor that electron $y$ does not. In $A$ we just have a kind of \ita{weight} of energy which enables us to say that there is a certain number of electrons there (in the He atom, for instance, there are two electrons, but we cannot count them\footnote{By `counting', we mean the definition of a bijection between the collection of the two electrons and the von Neumann ordinal number $2 = \{0,1\}$. Really, to which electron should we attribute the number $0$?}), but never that this or that electron belongs to it.\footnote{Some authors think that since the cardinal of the collection is greater than one, the elements of the collection are necessarily distinct. The existence of quasi-set theory, to be mentioned below, shows that this hypothesis can be discussed.}  Interesting enough, there is a mathematics grounding in the \ita{theory of  quasi-sets} which makes the trick of separating the notions of identity and indiscernibility so that  the equivalence (\ref{indist}) is a theorem of the theory \cite[\S 7.2.6]{frekra06}. In this theory, we can speak in collections having a certain cardinal (termed its \ita{quasi-cardinal}) but without providing them an identity. So, to make the above informal language precise, we can say that the collection (quasi-set) of the electrons of the neutral atom $A$ is two, while the quasi-cardinal of the ion $A^+$ is one.
But this theory is not the story to be told for now (but see below). So, let us go back to the wells.

\section{Back to the wells}\label{back}
Let us consider again the wave functions $\psi_1$ and $\psi_2$ of two particles initially prepared in an entangled state and sent  each one to one of  two wells, which we call well 1 and well 2, and of course suppose that the experience can be performed (recall that infinite potential wells are idealizations). As said before, these wave functions describe the stationary states of the wells, but shall be read also as marking that there are  two particles, one in each  well. 

The question is: quantum mechanics says that there is no way to tell which particle is which due to the entangled state. But, in representing the two wells in the standard Hausdorff space and time setting, do the wells attribute identity to the particles? I don't think so, and let me explain why.  As said before, to have identity means to have an identity card, a something which enables us to distinguish the entity in other situations. Furthermore, this identity is \ita{extensional} in the sense explained earlier: the contexts change when \ita{different} (here this notion makes perfect sense) things are interchanged. Quantum particles don't have identity in this sense. Although trapped in the infinite wells, they have only what Toraldo di Francia has termed \ita{mock individuality}, an individuality (and, we could say, a `mock identity') that is lost as soon as the wells are open or when another similar particle is added to the well (if this was possible). And this of course cannot be associated with the idea of \ita{identity}. Truly, there is no identity card for quantum particles.\footnote{I shall not consider Bohmian quantum mechanics here, but it is not free of problems of different orders, as for instance in being counter-intuitive --- see the discussions in \cite{des}, mainly around page 19.} They are not individuals, yet can be isolated by trapping them for some time.

Some time ago I wrote a paper contesting the idea that Hans Dehmelt's positron Priscilla is an individual (or that it has identity) on the same grounds (\cite{kra11} including for references). Any positron could act as Priscilla as well, while no other person could substitute Donald Trump (I suppose that no one could say what he says and think as he thinks). Trump is supposed to be an individual,  he has identity, Priscilla surely is not, although both can (in principle) be \ita{isolated}, that is, taken as a `separate object'. Anyway, as I have argued in my paper, this is not enough to confer \ita{identity} to the positron (Trump is an individual by hypothesis, but his `identity', so as of whatever person, is an old philosophical problem --- we are here supposing that this is the case). 

But let us explore  the argumentation a little bit deeper. Suppose we aim at to describe the two particles at once, one in each well. Since they are `identical', we need to use (anti-)symmetric wave functions, something like (let us consider the anti-symmetric case) 
\begin{equation}\label{psi12}
\psi_{12}(a,b) = \frac{1}{\sqrt{2}}\Big(\psi_1(a) \psi_2(b) - \psi_1(b) \psi_2(a)\Big)
\end{equation}

\noindent in the Hilbert space $\mathcal{H}_1 \otimes \mathcal{H}_2$ where the sub-indexes name the wells and $a$ and $b$ `name' the particles. Important to recall that the use of (anti-)symmetric vectors intend precisely to make these labels not conferring identity to the particles. 
Thus, 

$$|\psi_{12}(a,b)|^2 = \frac{1}{2}\Big(|\psi_1(a) \psi_2(b)|^2 + |\psi_1(b) \psi_2(a)|^2 - 2 Re\Big\langle \psi_1(a) \psi_2(b) \Big|  \psi_1(b) \psi_2(a) \Big\rangle\Big).$$

Let us consider the interference term. We have that 

$$\Big\langle \psi_1(a) \psi_2(b) \Big|  \psi_1(b) \psi_2(a) \Big\rangle_{\mathcal{H}_1 \otimes \mathcal{H}_2} = \langle \psi_1(a)|\psi_1(b)\rangle_{\mathcal{H}_1} \cdot \langle \psi_2(b) | \psi_2(a) \rangle_{\mathcal{H}_2} = 0$$

\noindent since it is \ita{supposed} that $\psi_1(b) = \psi_2(a) = 0$ for $b$ is out of well 1 and $a$ is out of well 2. This supposition makes the interference term is null and this is interpreted as saying that there is no interaction between the particles. 


The most we can do is to identity the particle (calling it $a$) with well 1, but it would be indifferent for all physical purposes if in well 1 the particle was $b$ instead, if such names would make physical sense (but note that they make \ita{mathematical} sense). The particles don't have identity in the above sense and, without identity, the most we can say is that we have two wells (mathematically seen as disjoined sets) with cardinal 1 each, and that their elements are indistinguishable. 
So far, so good. Some further remarks are in order, for physics need to enter the discussion.

As said before, infinite potential wells are idealizations. They don't exist and cannot be constructed. The most we can say is that we have two very hight wells, which despite the great potential involved, do not avoid tunneling and perhaps nonlocal interactions between the wells. So we \ita{really} cannot say that the particles are in fact non interacting. But let us continue with the mathematical description as a \ita{Gedankenexperiment}. 

Since the interference term is \ita{assumed} to be null, the state (\ref{psi12}) is separable, so 
the probabilities, then, result from the probabilities of the wells separately. That is, we have
\begin{equation}\label{pp}
|\psi_{12}(a,b)|^2 = \frac{1}{2}\Big(|\psi_1(a) \psi_2(b)|^2\Big) + \frac{1}{2}\Big(|\psi_1(b) \psi_2(a)|^2\Big).
\end{equation}

What does this equation mean? Let us provide it an interpretation.

\paragraph{Interpretation}
Let's read the second member of this equality as indicating that we have the sum of the probability of particle $a$ be in well 1 \ita{and} particle $b$ in well 2 (let us call $\alpha$ this case), and the probability of particle $b$ be in well 1 \ita{and} particle $a$ be in well 2 (call $\beta$ this assumption). So, the second member is speaking about probabilities. But probabilities of what? The first member of the equation gives the answer.

So, here we are: what about the first member of (\ref{pp})? To give it an interpretation, we recall that one of the postulates of the calculus of probability says that $P(\alpha \cup \beta) = P(\alpha) + P(\beta)$ if $\alpha$ and $\beta$ are mutually exclusive events, which is the case, once we have assumed that the interference term is null. So, the first member of (\ref{pp}) can be read as indicating that we have a probability of particle $a$ be in well 1 and particle $b$ in well 2, \ita{or} particle $b$ be in well 1 and particle $a$ be in well 2. In reading the equality this way, it results that there is no way to know  which particle is which! It is a quantum fact (entailed by quantum mechanics) that in such a situation, described by states (\ref{psi12}) or (\ref{pp}), we can no longer say which particle is in which well.  We name the particles \ita{after} having them in the wells, not before; so, all we have are \ita{mock names}, which make no sense after they leave the wells.

Summing up, there is no identity (in the sense described above of an identity card) attributed to the particles, for any permutation between them conduces to exactly the same results in the same sense that whatever trapped positron would be Dehmelt's Priscilla as well \cite{kra11}. They are not individuals, which import for distinguishing permutations. I recall once more that one thing is to individuate an object, put it in isolation from others. This is an epistemological notion. Another one is to confer it an identity; as Hume has said a long time ago, "One single object conveys the idea of unity, not that of identity" \cite[p.200]{hum86}. Dehmelt, Haroche, and Wineland's trapped quanta are not individuals having identities, yet individualized (concerning Dehmelt, see the references in \cite{kra11}; as for Haroche and Wineland, see \cite{royal12}). As for the differences among the notions of identity, individuality, and individuation, see \cite{kraare18}.

In other words, the physical interpretation does not enable us to identify the particles, confer identity to them, even if metamathematically they can be separated and provided them with a \ita{mathematical identity} by their classical space location. Thus, we have here a typical case where the language of the mathematics of the theory and the language of physics do not conform each other. 
And, as far as a clear analysis of any logical system also comprises a semantics,\footnote{Let us exemplify this talk with a case in logic. We can say that intuitionistic logic, or better Brouwer-Heyting logic, can be obtained from classical logic by just dropping the excluded middle. Rigorously speaking, this is a mistake, and can have a clear meaning only from a purely syntactic point of view, for classical logic and intuitionistic logic have different semantics, and a logic can make sense only within a semantic consideration --- see \cite{cosbuebez95}.} a physical theory cannot rest on pure mathematics. And in the quantum case, when this happens, there appears to be a conflict between the mathematics used (NST) and the physics.

A warning remark: the sense of the word `semantics' is not clear in general. It has different meanings in different contexts, so as the notion of \ita{interpretation} os a physical theory, mainly in what respects quantum mechanics. As put in \cite{cosbuebez95}, when first proposed, the term `semantics' was used to denote that part of logic concerning the determination of the meanings of the well formed expressions (formulas). "More recently, however, faced with an enormous variety of alternative meanings, it is no longer possible to specify an exact sense of this word" (ibid.). Concerning physical sciences, and in particular NQM, we are not speaking (here) of \ita{formal} semantics, but in informal ones, more or less in the sense of Omnès, just in giving physical sense to the mathematical expressions, but senses which are suitable for doing quantum physics. 


\section{The (supposed) right  answer}
Going to the conclusions, I shall propose the following. Quantum mechanics enables us to \ita{isolate} a quantum system, say by trapping it, give it a \ita{mock individuation}, taken here as an epistemological notion of we being able to look to the trap and guess that there is a particle there. But  the traps, so as the wells, do not confer it an identity card, but just a \ita{mock identity},\footnote{In the above sense of numerical identity, for other kinds of identity, such as \ita{relative identity} are not `identity' strictly speaking, but forms of indistinguishability.} something that is lost as soon as it merges with others of similar species. 

Furthermore, we should recall that from the physical point of view, the particles (seen either as point particles in non-relativistic quantum mechanics or in quantum field theories) don't have well defined positions in the standard sense, but just  position-states, so they cannot be precisely located in an specific point, but in a region, and even so with a given probability. What we have, before a measurement, is a superposed state of all possible positions of the particle. Even after a measurement, that we call `particle' becomes confined not to a point, but to a region of space (recall our above discussion about the de Broglie wavelength, and see the Appendix). Problems will arise if the uncertainty about the particle's position is zero (due to Heinseberg's Uncertainty Principle).  Mathematically, the position operator doesn't have eingenfunctions in the standard sense. The use of distributions in a rigged Hilbert space (Dirac's delta-function) just provides an approximation for the positions, and in reality we should go to quantum field theories, where space and time become (Minkowskian) spacetime but the problem of identity doesn't disappear (Wilczek, he again, have answered the question  "You have stated what you believe to be the single most profound result of quantum field theory. Can you repeat for the court what that is?" by saying "That two electrons are indistinguishable" \cite{gef15}). To see why quantum systems cannot have definite positions in the classical sense, perhaps the best thing is to read the quotation below, taken from Viktor Toth's answer to "Do subatomic particles have solid surfaces?" on \ita{Quora} \cite{tot16}, which helps in enlighten the point (my emphases):

\begin{quote}
Question: Do subatomic particles have solid surfaces?

Answer: 
"No. Subatomic particles are not like anything you experience in the classical world. In particular, they are not miniature cannonballs or planets or whatever.

In a quantum particle theory, elementary particles are point-like. However, \textit{these points do not usually have a classically defined location}, not unless they interact with other things (e.g., an instrument) that confines them to a location. \textit{This is why, instead of a position coordinate, the particle's location is described by its wave-function, which basically provides a probability field, assigning to each location of space a probability of finding the particle at that location.}"



\end{quote}

In short, contrary to Quine, we can say that quantum mechanics presents us entities (taking this as a very general term) with no identity, and quasi-set theory is a mathematics able to deal with them, for its quantifiers can range over collections of absolutely indiscernible non-individuals. So, these entities can be values of bound variables \cite{qui48}.

\section{Foundations}
What should we do when we realize that the theory does not conform with the (perhaps \ita{Gedanken}) experiments? What did Einstein  when he realized that the Newtonian space and time did not conform with the proposed ideas of the special relativity? What did von Neumann and Birkhoff when they realized that the `logic' of quantum mechanics didn't conform to classical logic? Cases like these could be mentioned to exhaustion. Well, Einstein proposed to change the notions of space and time, and we gained the sole notion of spacetime (Minkowski); von Neumann and Birkhoff proposed that quantum logic should be non-distributive, hence not classical. The reasons they had are of deep nature, and we all know about the consequences.

What to say about the above discussion involving classical space and time and the very nature of quantum particles, at least in what concerns their properties mentioned in the previous sections? It seems to me that in ignoring physics and paying attention just to the underlying mathematics, for example in saying that the standard Newtonian space and time framework provides identity to quantum objects when they are separated in space (let me for a moment call this thesis Thesis of Identity, TI), conflicts with the physics, for as we have seen that from the physical point of view there are no perfect separated quantum objects. To assume TI without considering physics is similar to consider a logic without paying attention to its semantic aspects. In the same vein that a logic comprises not only its syntactical aspects, but also its semantics, a physical theory cannot be grounded only in its mathematics; physics, that is, physical suppositions, need also to be considered. 

This was what motivated  above mentioned authors (Einstein, von Neumann and Birkhoff) to depart from classical frameworks and propose radically different bases. The same seems to be happing here. Newtonian space and time and the TI thesis, at least to me, conflict with quantum mechanics. I still don't know what kind of space and time or spacetime should be the right one, perhaps a Minkowskian is enough, but I still don't know. But it seems clear to me that there is a contradiction, at least at the metalevel, between  quantum mechanics and its underlying mathematics also in this respect (for another kind of discrepancies, see \cite{frekra06}, \cite{arekra14} where the discrepancies of the classical mathematical and logic notion of individual conflicts with the non-individuality of quantum objects).

The search for a right foundational mathematical basis for QM is still an open problem.

\section{Appendix 1 :  \\ How space and time enter the quantum schema}
We shall use the Hilbert space formalism, common in most philosophical discussions. From a technical point of view, we can assume that we shall be working with the resources of ZFC. 
So, let us introduce the following definition (more details in \cite{kraare16}), which follows the style presented in \cite[Chap.12]{sup57}:\footnote{Here we introduce a set $\mcal{S}$ for the quantum systems being considered, which is not useful in the standard presentations. The aim is to question either such a collection can be considered as a \ita{set} of standard set theories, since the quantum systems may be indiscernible. But this point will be not discussed here --- see \cite{arekra14}.} 

\begin{dfn} {\rm A non relativistic quantum mechanics  is a 5-tuple of the form $$\mathcal{Q} = \langle  S, \{\mathcal{H}_i\}, \{\hat{A}_{ij}\}, \{\hat{U}_{ik}\}, \mathcal{B}(\mathbb{R})\rangle, \; \mathrm{with} \; i \in I, j \in J, k \in K$$ \noindent where:
\begin{enumerate}
\item $S$ is a collection whose elements are called \ita{physical objects}, or \ita{physical systems}. Here is a novelty in our approach. The standard formalism doesn't speak of the quantum systems directly, making reference only to their states and observables. The talk about the systems is made in the metalanguage. Here we are introducing the systems in the structure. Since they can be indiscernible in a strong sense, $S$ would not be a set, but a quasi-set. But, if we aim at to consider this last option, we need to change ZFC by quasi-set theory.
\item  $\{H_i\}$ is a collection of mathematical structures, namely, complex separable Hilbert spaces whose cardinality is defined by the particular application of the theory. 
\item  $\{\hat{A}_{ij}\}$ is a collection of self-adjunct (or Hermitian) operators over a particular Hilbert space $H_i$.
\item $\{U_{ik}\}$ is a collection of unitary operators over a particular Hilbert space $H_i$\item  $\mathcal{B}(\mathbb{R})$ is the collection of Borel sets\index{Borel sets} over the set of real numbers.
\end{enumerate}}
\end{dfn}

In order to connect the formalism with experience, we construct a mathematical framework for representing experience. Important to remark that this is another theoretical (abstract) construction: there is no connection, out of the informal one, of the formalism with reality \ita{per se}. In order to do it, we need to elaborate reality, turning it a mathematical construct too. So, to each quantum system $s \in \mcal{S}$ we associate a 4-tuple $$\sigma = \langle \mathbb{E}^4, \psi(\mathbf{x}, t), \Delta, P\rangle,$$ \noindent where $\mathbb{E}^4$ is the Newtonian spacetime \cite[chap.17]{pen05}, where each point is denoted by a 4-tuple $\langle \mathbf{x}, t\rangle$ where $\mathbf{x} = \langle x, y, z \rangle$ denote the coordinates of the system and $t$ is a parameter representing time, $\psi(\mathbf{x},t)$ is a function over $\mathbb{E}^4$ called the \ita{wave function} of the system, $\Delta \in \mathcal{B}(\mathbb{R})$ is a Borelian, and $P$ is a function defined, for some $i$ (determined by the physical system $s$), in $\mathcal{H}_i \times \{\hat{A}_{ij}\} \times \mathcal{B}(\mathbb{R})$ and assuming values in $[0,1]$, so that the value $P(\psi, \hat{A},  \Delta) \in [0,1]$ is the probability that the measurement of the observable $A$ (represented by the self-adjunct operator $\hat{A}$) for the system in the state $\psi(\mathbf{x},t)$ lies in the Borelian set $\Delta$.\footnote{Of course if we have a system with $n$ elements, the dimension of the space must be $3n$ (here, roughly, $\mathbb{E}^4 = \mathbb{R}^3 \times \mathbb{R}$, and in the case with $n$ systems, we shall have $\mathbb{R}^{3n} \times \mathbb{R}$).}

Summing up: using standard mathematics, and classical space and time setting, this is all we can do, and it works! Physics can be made within such a framework, but philosophically we can raise questions such as those posed above.  Summing up, I can give an answer to Omnès question about the consistence of the two languages: NO, in general they are not consistent one another.

\section{Appendix 2 :  \\ A non Hausdorff topology for QM?}
In this Appendix, we advance an idea to be developed further, but which can complete the above discussion. The intention here is (I hope) to get some feedback. We shall be supposing the theory of quasi-sets  (qsets) without further details \cite{arekra14}, \cite{frekra06}, \cite{kraare18}. Informally speaking, qsets may comprise absolutely indiscernible elements, entities to which the standard sense of identity ascribed by classical logic and set theory does not apply. In this theory, we can speak of indistinguishable but not identical objects. The qsets may have a cardinal, termed its \ita{quasi-cardinal} even if its elements cannot be discerned form one another (this demands explanation, for it is usually --- and wrongly, according to the mentioned works --- supposed that once a collection has a cardinal greater than one, its elements are distinguishable). So, let us go.

Consider a qset $A$ whose elements are all indiscernible one each other (we write $x \equiv y$ for all $x, y \in A$). Let $B$ be a proper subqset of $A$. First, to avoid misunderstandings, let us explain why sentences such as "there is an element of $A$ that belong to $B$" doesn't imply that \ita{all} elements of $A$ belong to $B$, as it could be supposed since the elements of $A$ are indistinguishable. It is in this sense that we can speak of a `proper' subqset of $A$. The explanation is in term of cardinals, the \ita{quasi-cardinals} of the qsets. When a qset has a quasi-cardinal, or just q-cardinal for short (not always a qset has a q-cardinal), then this q-cardinal is a cardinal in the standard sense, defined in the \ita{classical part} of the theory (which is a model of ZFU, the Zermelo-Fraenkel set theory with \ita{Urelemente}). Furthermore, it is postulated that for every qset $A$ whose q-cardinal is $\lambda$, there are subqsets of $A$ whose q-cardinals are $\eta$ for every $\eta \leq \lambda$. So, in our hypothesis, once we have assumed that $A$ has a not zero q-cardinal, then it makes sense in the theory to speak of a `proper' subqset $B$, whose q-cardinal is less that the q-cardinal of $A$. So, although the elements of $A$ are indiscernible, the theory is consistent with the affirmative that not every element of $A$ is an element of $B$. 

Now, let us define the \ita{cloud} of $B$ as being the qset of the elements of $A$ that \ita{could} be elements of $B$ (we could call them the \ita{potential} elements of $B$). The formal definition says that $Cl_A(B)$, the cloud of $B$ relative to $A$, is the qset of all elements of $A$ which are indistinguishable from some element of $B$.\footnote{This is something accepted in quantum physics. What electrons are in the level $2p$ of a Sodium atom? There is no sense in this question. There are not \ita{electrons there}, but a stable number of them (whatever thing they are)  \ita{visiting} that shell in each instant of time.} In the supposed case, the cloud of $B$ would be the whole $A$ since its elements are indiscernible, so any element of $A$ \ita{could be} in $B$. Now let us consider another subqset $C \subseteq A$ \ita{disjoined} from $B$. Depending on the q-cardinals of $A$, $B$ and $C$, this is possible to suppose. For instance, suppose that the q-cardinal of $A$ is 6 (take $A$ as formed by the six electrons of the level $2p$ of a Sodium atom). Then the axioms of the theory are compatible with the intuitive idea (which in standard set theories is a theorem) that there are three disjoined subqsets of $A$ with q-cardinals 2 each. The interesting thing is that we have no means to know \ita{which} elements belong to each subqset, in the same vein that we can talk of two of the six electrons in the $2p$ shell but have no means to identify them. It is in this sense that we can say that the three subqsets are \ita{disjoined}. But their clouds relative to the whole collection coincide with $A$. 

So, let us go back to our example. We had two disjoined subqsets of $A$, but their clouds relative to $A$ intercept (in the given example, they coincide with $A$), so they are not disjoined. If we refer to $B$ and $C$ as the \ita{core} qsets, then there is a sense in saying that they do not have common elements but that their clouds do have and, depending on the characteristics of $A$, will always intercept. 

Let us consider now the \ita{algebra of clouds}. By this we mean, as in standard mathematics, the consideration of the clouds of the subqsets of some given qset (not necessarily with all elements indiscernible) with the usual mathematical operations (which have their counterpart in the theory of qsets). If we were considering sets, the algebra, as is known, would be a Boolean algebra. But here, as shown in \cite{naskrafei11}, this is not the case. The algebraic counterpart of the clouds is a non distributive lattice we have termed \ita{lattice of indiscernibles}. It resembles in much an orthomodular lattice, typical of quantum mechanics. But this is not what we intend to do here. 

What we wish to say is that by considering qsets and their clouds, we get a (quasi-)set theoretical framework where for any \ita{point} (element of a qset) we can consider a \ita{core} to which it belongs to which is disjoined from another \ita{core} of some `other' element of the qset but so that their clouds intercept.\footnote{We can use the above analogy that the electrons are just "visiting" the shells of an atom and say that the elements of $A$ may be "visiting" the core qsets, but also that an elements cannot be in two places at once.} So, the cores can be though as the two wells of sections (\ref{idd}) and (\ref{back}). Since their clouds intercept, we can say that the elements are never \ita{completely isolated}.  The talk of `other' element is, again, just a way of speech. All can be expressed in terms of q-cardinals. 

The advantage of such a qset-theoretical approach is that the elements can be considered as non-individuals, as entities without a Principle of Individuality, without identity, contrary to what occur in a standard set theory, where of course we can do similar things, for instance by considering either fuzzy sets  or \ita{quasets} \cite{daltor93} (see also \cite[Sect.7.4]{frekra06}). But, in these cases, the objects would have identity, are individuals, contrary to our preferred metaphysics of seeing then as entities devoid of identity conditions. 

The next step is to define a topology in qset theoretical terms. The definition is standard. We say that a qset $\Xi = \langle X, \tau \rangle$ is a q-topological space if $X$ is a qset and $\tau$ is a family of clouds of subqsets of $X$ satisfying the following conditions:
\begin{enumerate}
\item  $\emptyset$ and $X$ belong to $\tau$
\item  $A \cap B \in \tau$ for every clouds $A$ and $B$ in $\tau$
\item For any family of clouds of $\tau$, their union also belongs to $\tau$
\end{enumerate}

The definition is standard,  but the consequences in taking it within quasi-set theory are not. 
So, where is the difference to a standard topology? The answer is that the clouds may ever intercept (depending of the chosen qsets). In this sense, the space is not Hausdorff and, consequently, give that we represent two similar quantum objects within such a frameowrk, we cannot say that there are disjoined open balls centered in these points which do not intercept, since the `balls' are taken as clouds of some qset. It seems that this non Hausdorff topology fits better what NQM dictates. But a qset topology is still something to be pursued. 

\section*{Acknowledgements} I would like to thank Acácio José de Barros, Frederico Firmo de Souza Cruz and Graciela Domenech for kind discussions on the first draft  of this paper. The faults that remain are mine.


\begin{thebibliography}{999}
\bibitem[AreKra.14]{arekra14} Arenhart, J. R. B. \& Krause, D. (2014), From primitive identity to the non-individuality of quantum objects. \ita{Studies in History and Philosophy of Modern Physics}  46: 273-282.
\bibitem[BirVon.36]{birvon36} Birkhoff, G. \& von Neumann, J. (1936), The logic of quantum mechanics, \ita{Annals of Mathematics}, vol. 37, pp. 823?843.
\bibitem[BokJae.10]{bokjae10} Boukulich, A. and Jaeger, G. (2010), \ita{Philosophy of Quantum Information and Entanglement}. Cambridge: Cambridge Un. Press. 
\bibitem[Bub.17]{bub17} Bub, J., Quantum Entanglement and Information, \ita{The Stanford Encyclopedia of Philosophy (Spring 2017 Edition)}, Edward N. Zalta (ed.), URL = <https://plato.stanford.edu/archives/spr2017/entries/qt-entangle/>.
\bibitem[CosBueBez.95]{cosbuebez95} da Costa, N. C. A., Bueno, O. \& Béziau, J. -Y. (1995), What is semantics? A note on a huge question. \ita{Sorites} (November): 43-47. 
\bibitem[DalTor.93]{daltor93} Dalla Chiara, M. L. and Toraldo di Francia, G. (1993), Individuals, kinds and names in physics, in Corsi, G.\
\textit{et al.} (eds.), \textit{Bridging the gap: philosophy,
mathematics, physics}, Kluwer Ac.\ Pu., pp. 261-83.
\bibitem[DalGiuGre.04]{dalgiugre04} Dalla Chiara, M.L., Giuntini, R. \& Greechie, R. (2004), \ita{Reasoning in Quantum Theory: Sharp and Unsharp Quantum Logics}. Dordrecht/Boston/London: Kluwer Ac. Pu. (Trends in Logic - \ita{Studia Logica Library} Vol. 22).
\bibitem[DesZwi.06]{des} D'Espagnat, B. \& Zwirn, H. (eds.), \ita{The Quantum World: Philosophical Debates on Quantum Physics}. Springer (The Frontiers Collection).
\bibitem[FreKra.06]{frekra06} French, S. \& Krause, D. (2006), \ita{Identity in Physics: A Historical, Philosophical, and Formal Analysis}. Oxford: Oxford Un. Press. 
\bibitem[Gef.15]{gef15} Gefter, A. (2015), Quantum mechanics is putting human identity on trial. In \href{https://goo.gl/fHESGq}{Nautilus}.
\bibitem[Gri.95]{gri95} Griffiths, D. J. (1995), \ita{Introduction to Quantum Mechanics}. New Jersey: Prentice Hall.
\bibitem[Hum.86]{hum86} Hume, D. (1986] [1739], \ita{A Treatise on Human Nature} (L. A. Selby-Bigge ed.). Oxford: Oxford Un. Press.
\bibitem[Jam.66]{jam66} Jammer, M. (1966), \ita{The Conceptual Development of Quantum Mechanics}. New York: McGraw-Hill.
\bibitem[LanLif.91]{lanlif91} Landau, L. D. \& Lifshitz, E. M. (1991), \ita{Quantum Mechanics: Non-Relativistic Theory}. 3rd. ed., Pergamon Press.
\bibitem[Kra.11]{kra11} Krause, D. (2011), Is Priscilla, the trapped positron, an individual? Quantum physics, the use of names, and individuation. \ita{Arbor} 187 (747):  61-66. 
\bibitem[KraAre.16]{kraare16} Krause, D. \& Arenhart, J. R. B. (2016), \ita{The Logical Foundations of Scientific Theories: Languages, Structures, and Models}. London: Routledge.
\bibitem[KraAre.18]{kraare18}  Krause, D.  \& Arenhart, J. R. B. (2018), Quantum non-individuality: background concepts and possibilities. To appear in S. Wuppuluri \& F. A. Doria (eds.), \ita{Map and Territory, Exploring the Foundations of Science, Thought and Reality}. Springer Frontier Series, January 2018.
\bibitem[Omn.99]{omn99} Omnés, R. 1999,  Recent Advances in the Consistency o f Interpretation. In 
D. Greenberger eta/. (eds.), 
\ita{Epistemological and Experimental Perspectives on Quantum Physics} (Vienna Circle Institute Yearbook 1999). Dordrecht:  Kluwer Academic Publishers, 103-11.
\bibitem[NasKraFei.11]{naskrafei11} Nascimento, M. C., Krause, D. \& Feitosa, H. de A 2011. The quasi-lattice of indiscernible elements. \ita{Studia Logica} 97: 101-128.
\bibitem[Pen.05]{pen05} Penrose, R. (2005), \ita{The Road to Reality: A Complete Guide to the Laws of the Universe}. New York: Alfred A. Knoff.
\bibitem[Pos.63]{pos63} Post, H. (1963), Individuality and Physics, \ita{The Listener}, 10th October 1963, pp. 534-537. Reprinted in \ita{Vedanta for East and West} 132, 1973, pp. 14-22.
\bibitem[Roy.12]{royal12} Royal Swedish Academy of Sciences (2012), Scientific Background on the Nobel Prize in Physics 2012: measuring and manipulating individual quantum systems. {\tt{https://goo.gl/1gT187}}
\bibitem[Rov.17]{rov17} Rovelli, C. (2017), "Space is blue and birds fly through it", \href{http://philsci-archive.pitt.edu/14180/1/Royal2017.pdf}{Phil.Sci.}
\bibitem[Qui.48]{qui48} Quine, W. O. (1948), On what there is. \ita{The Review of Metaphysics} 2 (1): 21-38.
\bibitem[Ste.67]{ste67} Stein, H. (1967), Newtonian space-time. \ita{Texas Quarterly} 10: 174-200.
\bibitem[Sty.01]{sty01} Styer, D. F. et al. (2001), Nine formulations of quantum mechanics. \ita{Am. J. Phys.} 70 (3): 288-97.
\bibitem[Sup.59]{sup57} Suppes, P. (1957), \ita{Introduction to Logic}. New York and Toronto: Van Nostrand.
\bibitem[Tel.98]{tel98} Teller, P. (1998), Quantum mechanics and haecceities. In Castellani, E. (ed.), \ita{Interpreting Bodies: Classical and Quantum Objects in Modern Physics}. Princeton: Princeton Un. Press, 115-141.
\bibitem[Tot.16]{tot16} Toth, V. T. (2016), Do subatomic particles have solid surfaces?, \href{https://goo.gl/MdAzKc}{Quora}
\bibitem[Wea.17]{wea17} Weatherall, J. O. (2017), Category theory and the foundations of classical space-time theories. In Laundry, E. (ed.), \ita{Categories for the Working Philosopher}. Oxford: Oxford Un. Press, pp. 329-47.

\end{thebibliography}
\end{document}